%% file: cp_bpmds13.tex
\documentclass[a4paper,lnbip]{svmultln}
\usepackage{graphicx}
\usepackage{float}

\graphicspath{{img/}}

\usepackage[%
final,%
usepdfnote,%
showoldrev,%
marknewrev%
]{todos} 
\usepackage{color}
\input{rgb}

\begin{document}
\mainmatter
 
\title{Change Patterns in Use: A Critical Evaluation\thanks{This
research is supported by Austrian Science Fund (FWF): P23699-N23. The final
publication is available at Springer via
http://dx.doi.org/10.1007/978-3-642-38484-4\_19}}
\titlerunning{Change Patterns in Use: A Critical Evaluation} 
\author{Barbara Weber\inst{1} \and Jakob Pinggera\inst{1} \and Victoria
Torres\inst{2} \and Manfred Reichert\inst{3}}
\authorrunning{Weber et al.}
 
\institute{University of Innsbruck, Austria\\
\email{{barbara.weber, jakob.pinggera}@uibk.ac.at}  
\and Universitat Polit\`ecnica de Val\`encia, Spain\\
\email{vtorres@pros.upv.es}
\and University of Ulm, Germany\\
\email{manfred.reichert@uni-ulm.de}
}

\maketitle

\begin{abstract}
Process model quality has been an area of considerable research efforts. In this
context, the correctness-by-construction principle of change patterns provides
promising perspectives. However, using change patterns for model creation imposes
a more structured way of modeling. While the process of process modeling (PPM)
based on change primitives has been investigated, little is known about this
process based on change patterns. To obtain a better understanding of the PPM
when using change patterns, the arising challenges, and the subjective
perceptions of process designers, we conduct an exploratory study. The results indicate that
process designers face little problems as long as control-flow is simple, but
have considerable problems with the usage of change patterns when complex, nested
models have to be created. Finally, we outline how effective tool support for
change patterns should be realized.

\end{abstract} 
 
\vspace{-1cm}
 
\input{introduction}

\input{backgrounds-vt}

\input{casestudy}

\input{results}

\input{discussion}
\input{relatedwork}

\input{summary}

\bibliographystyle{splncs}
\bibliography{literature}

\end{document}

%% file: rgb.tex
\definecolor{AliceBlue}{rgb}{0.94,0.97,1.00}
\definecolor{AntiqueWhite1}{rgb}{1.00,0.94,0.86}
\definecolor{AntiqueWhite2}{rgb}{0.93,0.87,0.80}
\definecolor{AntiqueWhite3}{rgb}{0.80,0.75,0.69}
\definecolor{AntiqueWhite4}{rgb}{0.55,0.51,0.47}
\definecolor{AntiqueWhite}{rgb}{0.98,0.92,0.84}
\definecolor{BlanchedAlmond}{rgb}{1.00,0.92,0.80}
\definecolor{BlueViolet}{rgb}{0.54,0.17,0.89}
\definecolor{CadetBlue1}{rgb}{0.60,0.96,1.00}
\definecolor{CadetBlue2}{rgb}{0.56,0.90,0.93}
\definecolor{CadetBlue3}{rgb}{0.48,0.77,0.80}
\definecolor{CadetBlue4}{rgb}{0.33,0.53,0.55}
\definecolor{CadetBlue}{rgb}{0.37,0.62,0.63}
\definecolor{CornflowerBlue}{rgb}{0.39,0.58,0.93}
\definecolor{DarkBlue}{rgb}{0.00,0.00,0.55}
\definecolor{DarkCyan}{rgb}{0.00,0.55,0.55}
\definecolor{DarkGoldenrod1}{rgb}{1.00,0.73,0.06}
\definecolor{DarkGoldenrod2}{rgb}{0.93,0.68,0.05}
\definecolor{DarkGoldenrod3}{rgb}{0.80,0.58,0.05}
\definecolor{DarkGoldenrod4}{rgb}{0.55,0.40,0.03}
\definecolor{DarkGoldenrod}{rgb}{0.72,0.53,0.04}
\definecolor{DarkGray}{rgb}{0.66,0.66,0.66}
\definecolor{DarkGreen}{rgb}{0.00,0.39,0.00}
\definecolor{DarkGrey}{rgb}{0.66,0.66,0.66}
\definecolor{DarkKhaki}{rgb}{0.74,0.72,0.42}
\definecolor{DarkMagenta}{rgb}{0.55,0.00,0.55}
\definecolor{DarkOliveGreen1}{rgb}{0.79,1.00,0.44}
\definecolor{DarkOliveGreen2}{rgb}{0.74,0.93,0.41}
\definecolor{DarkOliveGreen3}{rgb}{0.64,0.80,0.35}
\definecolor{DarkOliveGreen4}{rgb}{0.43,0.55,0.24}
\definecolor{DarkOliveGreen}{rgb}{0.33,0.42,0.18}
\definecolor{DarkOrange1}{rgb}{1.00,0.50,0.00}
\definecolor{DarkOrange2}{rgb}{0.93,0.46,0.00}
\definecolor{DarkOrange3}{rgb}{0.80,0.40,0.00}
\definecolor{DarkOrange4}{rgb}{0.55,0.27,0.00}
\definecolor{DarkOrange}{rgb}{1.00,0.55,0.00}
\definecolor{DarkOrchid1}{rgb}{0.75,0.24,1.00}
\definecolor{DarkOrchid2}{rgb}{0.70,0.23,0.93}
\definecolor{DarkOrchid3}{rgb}{0.60,0.20,0.80}
\definecolor{DarkOrchid4}{rgb}{0.41,0.13,0.55}
\definecolor{DarkOrchid}{rgb}{0.60,0.20,0.80}
\definecolor{DarkRed}{rgb}{0.55,0.00,0.00}
\definecolor{DarkSalmon}{rgb}{0.91,0.59,0.48}
\definecolor{DarkSeaGreen1}{rgb}{0.76,1.00,0.76}
\definecolor{DarkSeaGreen2}{rgb}{0.71,0.93,0.71}
\definecolor{DarkSeaGreen3}{rgb}{0.61,0.80,0.61}
\definecolor{DarkSeaGreen4}{rgb}{0.41,0.55,0.41}
\definecolor{DarkSeaGreen}{rgb}{0.56,0.74,0.56}
\definecolor{DarkSlateBlue}{rgb}{0.28,0.24,0.55}
\definecolor{DarkSlateGray1}{rgb}{0.59,1.00,1.00}
\definecolor{DarkSlateGray2}{rgb}{0.55,0.93,0.93}
\definecolor{DarkSlateGray3}{rgb}{0.47,0.80,0.80}
\definecolor{DarkSlateGray4}{rgb}{0.32,0.55,0.55}
\definecolor{DarkSlateGray}{rgb}{0.18,0.31,0.31}
\definecolor{DarkSlateGrey}{rgb}{0.18,0.31,0.31}
\definecolor{DarkTurquoise}{rgb}{0.00,0.81,0.82}
\definecolor{DarkViolet}{rgb}{0.58,0.00,0.83}
\definecolor{DeepPink1}{rgb}{1.00,0.08,0.58}
\definecolor{DeepPink2}{rgb}{0.93,0.07,0.54}
\definecolor{DeepPink3}{rgb}{0.80,0.06,0.46}
\definecolor{DeepPink4}{rgb}{0.55,0.04,0.31}
\definecolor{DeepPink}{rgb}{1.00,0.08,0.58}
\definecolor{DeepSkyBlue1}{rgb}{0.00,0.75,1.00}
\definecolor{DeepSkyBlue2}{rgb}{0.00,0.70,0.93}
\definecolor{DeepSkyBlue3}{rgb}{0.00,0.60,0.80}
\definecolor{DeepSkyBlue4}{rgb}{0.00,0.41,0.55}
\definecolor{DeepSkyBlue}{rgb}{0.00,0.75,1.00}
\definecolor{DimGray}{rgb}{0.41,0.41,0.41}
\definecolor{DimGrey}{rgb}{0.41,0.41,0.41}
\definecolor{DodgerBlue1}{rgb}{0.12,0.56,1.00}
\definecolor{DodgerBlue2}{rgb}{0.11,0.53,0.93}
\definecolor{DodgerBlue3}{rgb}{0.09,0.45,0.80}
\definecolor{DodgerBlue4}{rgb}{0.06,0.31,0.55}
\definecolor{DodgerBlue}{rgb}{0.12,0.56,1.00}
\definecolor{FloralWhite}{rgb}{1.00,0.98,0.94}
\definecolor{ForestGreen}{rgb}{0.13,0.55,0.13}
\definecolor{GhostWhite}{rgb}{0.97,0.97,1.00}
\definecolor{GreenYellow}{rgb}{0.68,1.00,0.18}
\definecolor{HotPink1}{rgb}{1.00,0.43,0.71}
\definecolor{HotPink2}{rgb}{0.93,0.42,0.65}
\definecolor{HotPink3}{rgb}{0.80,0.38,0.56}
\definecolor{HotPink4}{rgb}{0.55,0.23,0.38}
\definecolor{HotPink}{rgb}{1.00,0.41,0.71}
\definecolor{IndianRed1}{rgb}{1.00,0.42,0.42}
\definecolor{IndianRed2}{rgb}{0.93,0.39,0.39}
\definecolor{IndianRed3}{rgb}{0.80,0.33,0.33}
\definecolor{IndianRed4}{rgb}{0.55,0.23,0.23}
\definecolor{IndianRed}{rgb}{0.80,0.36,0.36}
\definecolor{LavenderBlush1}{rgb}{1.00,0.94,0.96}
\definecolor{LavenderBlush2}{rgb}{0.93,0.88,0.90}
\definecolor{LavenderBlush3}{rgb}{0.80,0.76,0.77}
\definecolor{LavenderBlush4}{rgb}{0.55,0.51,0.53}
\definecolor{LavenderBlush}{rgb}{1.00,0.94,0.96}
\definecolor{LawnGreen}{rgb}{0.49,0.99,0.00}
\definecolor{LemonChiffon1}{rgb}{1.00,0.98,0.80}
\definecolor{LemonChiffon2}{rgb}{0.93,0.91,0.75}
\definecolor{LemonChiffon3}{rgb}{0.80,0.79,0.65}
\definecolor{LemonChiffon4}{rgb}{0.55,0.54,0.44}
\definecolor{LemonChiffon}{rgb}{1.00,0.98,0.80}
\definecolor{LightBlue1}{rgb}{0.75,0.94,1.00}
\definecolor{LightBlue2}{rgb}{0.70,0.87,0.93}
\definecolor{LightBlue3}{rgb}{0.60,0.75,0.80}
\definecolor{LightBlue4}{rgb}{0.41,0.51,0.55}
\definecolor{LightBlue}{rgb}{0.68,0.85,0.90}
\definecolor{LightCoral}{rgb}{0.94,0.50,0.50}
\definecolor{LightCyan1}{rgb}{0.88,1.00,1.00}
\definecolor{LightCyan2}{rgb}{0.82,0.93,0.93}
\definecolor{LightCyan3}{rgb}{0.71,0.80,0.80}
\definecolor{LightCyan4}{rgb}{0.48,0.55,0.55}
\definecolor{LightCyan}{rgb}{0.88,1.00,1.00}
\definecolor{LightGoldenrod1}{rgb}{1.00,0.93,0.55}
\definecolor{LightGoldenrod2}{rgb}{0.93,0.86,0.51}
\definecolor{LightGoldenrod3}{rgb}{0.80,0.75,0.44}
\definecolor{LightGoldenrod4}{rgb}{0.55,0.51,0.30}
\definecolor{LightGoldenrodYellow}{rgb}{0.98,0.98,0.82}
\definecolor{LightGoldenrod}{rgb}{0.93,0.87,0.51}
\definecolor{LightGray}{rgb}{0.83,0.83,0.83}
\definecolor{LightGreen}{rgb}{0.56,0.93,0.56}
\definecolor{LightGrey}{rgb}{0.83,0.83,0.83}
\definecolor{LightPink1}{rgb}{1.00,0.68,0.73}
\definecolor{LightPink2}{rgb}{0.93,0.64,0.68}
\definecolor{LightPink3}{rgb}{0.80,0.55,0.58}
\definecolor{LightPink4}{rgb}{0.55,0.37,0.40}
\definecolor{LightPink}{rgb}{1.00,0.71,0.76}
\definecolor{LightSalmon1}{rgb}{1.00,0.63,0.48}
\definecolor{LightSalmon2}{rgb}{0.93,0.58,0.45}
\definecolor{LightSalmon3}{rgb}{0.80,0.51,0.38}
\definecolor{LightSalmon4}{rgb}{0.55,0.34,0.26}
\definecolor{LightSalmon}{rgb}{1.00,0.63,0.48}
\definecolor{LightSeaGreen}{rgb}{0.13,0.70,0.67}
\definecolor{LightSkyBlue1}{rgb}{0.69,0.89,1.00}
\definecolor{LightSkyBlue2}{rgb}{0.64,0.83,0.93}
\definecolor{LightSkyBlue3}{rgb}{0.55,0.71,0.80}
\definecolor{LightSkyBlue4}{rgb}{0.38,0.48,0.55}
\definecolor{LightSkyBlue}{rgb}{0.53,0.81,0.98}
\definecolor{LightSlateBlue}{rgb}{0.52,0.44,1.00}
\definecolor{LightSlateGray}{rgb}{0.47,0.53,0.60}
\definecolor{LightSlateGrey}{rgb}{0.47,0.53,0.60}
\definecolor{LightSteelBlue1}{rgb}{0.79,0.88,1.00}
\definecolor{LightSteelBlue2}{rgb}{0.74,0.82,0.93}
\definecolor{LightSteelBlue3}{rgb}{0.64,0.71,0.80}
\definecolor{LightSteelBlue4}{rgb}{0.43,0.48,0.55}
\definecolor{LightSteelBlue}{rgb}{0.69,0.77,0.87}
\definecolor{LightYellow1}{rgb}{1.00,1.00,0.88}
\definecolor{LightYellow2}{rgb}{0.93,0.93,0.82}
\definecolor{LightYellow3}{rgb}{0.80,0.80,0.71}
\definecolor{LightYellow4}{rgb}{0.55,0.55,0.48}
\definecolor{LightYellow}{rgb}{1.00,1.00,0.88}
\definecolor{LimeGreen}{rgb}{0.20,0.80,0.20}
\definecolor{MediumAquamarine}{rgb}{0.40,0.80,0.67}
\definecolor{MediumBlue}{rgb}{0.00,0.00,0.80}
\definecolor{MediumOrchid1}{rgb}{0.88,0.40,1.00}
\definecolor{MediumOrchid2}{rgb}{0.82,0.37,0.93}
\definecolor{MediumOrchid3}{rgb}{0.71,0.32,0.80}
\definecolor{MediumOrchid4}{rgb}{0.48,0.22,0.55}
\definecolor{MediumOrchid}{rgb}{0.73,0.33,0.83}
\definecolor{MediumPurple1}{rgb}{0.67,0.51,1.00}
\definecolor{MediumPurple2}{rgb}{0.62,0.47,0.93}
\definecolor{MediumPurple3}{rgb}{0.54,0.41,0.80}
\definecolor{MediumPurple4}{rgb}{0.36,0.28,0.55}
\definecolor{MediumPurple}{rgb}{0.58,0.44,0.86}
\definecolor{MediumSeaGreen}{rgb}{0.24,0.70,0.44}
\definecolor{MediumSlateBlue}{rgb}{0.48,0.41,0.93}
\definecolor{MediumSpringGreen}{rgb}{0.00,0.98,0.60}
\definecolor{MediumTurquoise}{rgb}{0.28,0.82,0.80}
\definecolor{MediumVioletRed}{rgb}{0.78,0.08,0.52}
\definecolor{MidnightBlue}{rgb}{0.10,0.10,0.44}
\definecolor{MintCream}{rgb}{0.96,1.00,0.98}
\definecolor{MistyRose1}{rgb}{1.00,0.89,0.88}
\definecolor{MistyRose2}{rgb}{0.93,0.84,0.82}
\definecolor{MistyRose3}{rgb}{0.80,0.72,0.71}
\definecolor{MistyRose4}{rgb}{0.55,0.49,0.48}
\definecolor{MistyRose}{rgb}{1.00,0.89,0.88}
\definecolor{NavajoWhite1}{rgb}{1.00,0.87,0.68}
\definecolor{NavajoWhite2}{rgb}{0.93,0.81,0.63}
\definecolor{NavajoWhite3}{rgb}{0.80,0.70,0.55}
\definecolor{NavajoWhite4}{rgb}{0.55,0.47,0.37}
\definecolor{NavajoWhite}{rgb}{1.00,0.87,0.68}
\definecolor{NavyBlue}{rgb}{0.00,0.00,0.50}
\definecolor{OldLace}{rgb}{0.99,0.96,0.90}
\definecolor{OliveDrab1}{rgb}{0.75,1.00,0.24}
\definecolor{OliveDrab2}{rgb}{0.70,0.93,0.23}
\definecolor{OliveDrab3}{rgb}{0.60,0.80,0.20}
\definecolor{OliveDrab4}{rgb}{0.41,0.55,0.13}
\definecolor{OliveDrab}{rgb}{0.42,0.56,0.14}
\definecolor{OrangeRed1}{rgb}{1.00,0.27,0.00}
\definecolor{OrangeRed2}{rgb}{0.93,0.25,0.00}
\definecolor{OrangeRed3}{rgb}{0.80,0.22,0.00}
\definecolor{OrangeRed4}{rgb}{0.55,0.15,0.00}
\definecolor{OrangeRed}{rgb}{1.00,0.27,0.00}
\definecolor{PaleGoldenrod}{rgb}{0.93,0.91,0.67}
\definecolor{PaleGreen1}{rgb}{0.60,1.00,0.60}
\definecolor{PaleGreen2}{rgb}{0.56,0.93,0.56}
\definecolor{PaleGreen3}{rgb}{0.49,0.80,0.49}
\definecolor{PaleGreen4}{rgb}{0.33,0.55,0.33}
\definecolor{PaleGreen}{rgb}{0.60,0.98,0.60}
\definecolor{PaleTurquoise1}{rgb}{0.73,1.00,1.00}
\definecolor{PaleTurquoise2}{rgb}{0.68,0.93,0.93}
\definecolor{PaleTurquoise3}{rgb}{0.59,0.80,0.80}
\definecolor{PaleTurquoise4}{rgb}{0.40,0.55,0.55}
\definecolor{PaleTurquoise}{rgb}{0.69,0.93,0.93}
\definecolor{PaleVioletRed1}{rgb}{1.00,0.51,0.67}
\definecolor{PaleVioletRed2}{rgb}{0.93,0.47,0.62}
\definecolor{PaleVioletRed3}{rgb}{0.80,0.41,0.54}
\definecolor{PaleVioletRed4}{rgb}{0.55,0.28,0.36}
\definecolor{PaleVioletRed}{rgb}{0.86,0.44,0.58}
\definecolor{PapayaWhip}{rgb}{1.00,0.94,0.84}
\definecolor{PeachPuff1}{rgb}{1.00,0.85,0.73}
\definecolor{PeachPuff2}{rgb}{0.93,0.80,0.68}
\definecolor{PeachPuff3}{rgb}{0.80,0.69,0.58}
\definecolor{PeachPuff4}{rgb}{0.55,0.47,0.40}
\definecolor{PeachPuff}{rgb}{1.00,0.85,0.73}
\definecolor{PowderBlue}{rgb}{0.69,0.88,0.90}
\definecolor{RosyBrown1}{rgb}{1.00,0.76,0.76}
\definecolor{RosyBrown2}{rgb}{0.93,0.71,0.71}
\definecolor{RosyBrown3}{rgb}{0.80,0.61,0.61}
\definecolor{RosyBrown4}{rgb}{0.55,0.41,0.41}
\definecolor{RosyBrown}{rgb}{0.74,0.56,0.56}
\definecolor{RoyalBlue1}{rgb}{0.28,0.46,1.00}
\definecolor{RoyalBlue2}{rgb}{0.26,0.43,0.93}
\definecolor{RoyalBlue3}{rgb}{0.23,0.37,0.80}
\definecolor{RoyalBlue4}{rgb}{0.15,0.25,0.55}
\definecolor{RoyalBlue}{rgb}{0.25,0.41,0.88}
\definecolor{SaddleBrown}{rgb}{0.55,0.27,0.07}
\definecolor{SandyBrown}{rgb}{0.96,0.64,0.38}
\definecolor{SeaGreen1}{rgb}{0.33,1.00,0.62}
\definecolor{SeaGreen2}{rgb}{0.31,0.93,0.58}
\definecolor{SeaGreen3}{rgb}{0.26,0.80,0.50}
\definecolor{SeaGreen4}{rgb}{0.18,0.55,0.34}
\definecolor{SeaGreen}{rgb}{0.18,0.55,0.34}
\definecolor{SkyBlue1}{rgb}{0.53,0.81,1.00}
\definecolor{SkyBlue2}{rgb}{0.49,0.75,0.93}
\definecolor{SkyBlue3}{rgb}{0.42,0.65,0.80}
\definecolor{SkyBlue4}{rgb}{0.29,0.44,0.55}
\definecolor{SkyBlue}{rgb}{0.53,0.81,0.92}
\definecolor{SlateBlue1}{rgb}{0.51,0.44,1.00}
\definecolor{SlateBlue2}{rgb}{0.48,0.40,0.93}
\definecolor{SlateBlue3}{rgb}{0.41,0.35,0.80}
\definecolor{SlateBlue4}{rgb}{0.28,0.24,0.55}
\definecolor{SlateBlue}{rgb}{0.42,0.35,0.80}
\definecolor{SlateGray1}{rgb}{0.78,0.89,1.00}
\definecolor{SlateGray2}{rgb}{0.73,0.83,0.93}
\definecolor{SlateGray3}{rgb}{0.62,0.71,0.80}
\definecolor{SlateGray4}{rgb}{0.42,0.48,0.55}
\definecolor{SlateGray}{rgb}{0.44,0.50,0.56}
\definecolor{SlateGrey}{rgb}{0.44,0.50,0.56}
\definecolor{SpringGreen1}{rgb}{0.00,1.00,0.50}
\definecolor{SpringGreen2}{rgb}{0.00,0.93,0.46}
\definecolor{SpringGreen3}{rgb}{0.00,0.80,0.40}
\definecolor{SpringGreen4}{rgb}{0.00,0.55,0.27}
\definecolor{SpringGreen}{rgb}{0.00,1.00,0.50}
\definecolor{SteelBlue1}{rgb}{0.39,0.72,1.00}
\definecolor{SteelBlue2}{rgb}{0.36,0.67,0.93}
\definecolor{SteelBlue3}{rgb}{0.31,0.58,0.80}
\definecolor{SteelBlue4}{rgb}{0.21,0.39,0.55}
\definecolor{SteelBlue}{rgb}{0.27,0.51,0.71}
\definecolor{VioletRed1}{rgb}{1.00,0.24,0.59}
\definecolor{VioletRed2}{rgb}{0.93,0.23,0.55}
\definecolor{VioletRed3}{rgb}{0.80,0.20,0.47}
\definecolor{VioletRed4}{rgb}{0.55,0.13,0.32}
\definecolor{VioletRed}{rgb}{0.82,0.13,0.56}
\definecolor{WhiteSmoke}{rgb}{0.96,0.96,0.96}
\definecolor{YellowGreen}{rgb}{0.60,0.80,0.20}
\definecolor{aliceblue}{rgb}{0.94,0.97,1.00}
\definecolor{antiquewhite}{rgb}{0.98,0.92,0.84}
\definecolor{aquamarine1}{rgb}{0.50,1.00,0.83}
\definecolor{aquamarine2}{rgb}{0.46,0.93,0.78}
\definecolor{aquamarine3}{rgb}{0.40,0.80,0.67}
\definecolor{aquamarine4}{rgb}{0.27,0.55,0.45}
\definecolor{aquamarine}{rgb}{0.50,1.00,0.83}
\definecolor{azure1}{rgb}{0.94,1.00,1.00}
\definecolor{azure2}{rgb}{0.88,0.93,0.93}
\definecolor{azure3}{rgb}{0.76,0.80,0.80}
\definecolor{azure4}{rgb}{0.51,0.55,0.55}
\definecolor{azure}{rgb}{0.94,1.00,1.00}
\definecolor{beige}{rgb}{0.96,0.96,0.86}
\definecolor{bisque1}{rgb}{1.00,0.89,0.77}
\definecolor{bisque2}{rgb}{0.93,0.84,0.72}
\definecolor{bisque3}{rgb}{0.80,0.72,0.62}
\definecolor{bisque4}{rgb}{0.55,0.49,0.42}
\definecolor{bisque}{rgb}{1.00,0.89,0.77}
\definecolor{black}{rgb}{0.00,0.00,0.00}
\definecolor{blanchedalmond}{rgb}{1.00,0.92,0.80}
\definecolor{blue1}{rgb}{0.00,0.00,1.00}
\definecolor{blue2}{rgb}{0.00,0.00,0.93}
\definecolor{blue3}{rgb}{0.00,0.00,0.80}
\definecolor{blue4}{rgb}{0.00,0.00,0.55}
\definecolor{blueviolet}{rgb}{0.54,0.17,0.89}
\definecolor{blue}{rgb}{0.00,0.00,1.00}
\definecolor{brown1}{rgb}{1.00,0.25,0.25}
\definecolor{brown2}{rgb}{0.93,0.23,0.23}
\definecolor{brown3}{rgb}{0.80,0.20,0.20}
\definecolor{brown4}{rgb}{0.55,0.14,0.14}
\definecolor{brown}{rgb}{0.65,0.16,0.16}
\definecolor{burlywood1}{rgb}{1.00,0.83,0.61}
\definecolor{burlywood2}{rgb}{0.93,0.77,0.57}
\definecolor{burlywood3}{rgb}{0.80,0.67,0.49}
\definecolor{burlywood4}{rgb}{0.55,0.45,0.33}
\definecolor{burlywood}{rgb}{0.87,0.72,0.53}
\definecolor{cadetblue}{rgb}{0.37,0.62,0.63}
\definecolor{chartreuse1}{rgb}{0.50,1.00,0.00}
\definecolor{chartreuse2}{rgb}{0.46,0.93,0.00}
\definecolor{chartreuse3}{rgb}{0.40,0.80,0.00}
\definecolor{chartreuse4}{rgb}{0.27,0.55,0.00}
\definecolor{chartreuse}{rgb}{0.50,1.00,0.00}
\definecolor{chocolate1}{rgb}{1.00,0.50,0.14}
\definecolor{chocolate2}{rgb}{0.93,0.46,0.13}
\definecolor{chocolate3}{rgb}{0.80,0.40,0.11}
\definecolor{chocolate4}{rgb}{0.55,0.27,0.07}
\definecolor{chocolate}{rgb}{0.82,0.41,0.12}
\definecolor{coral1}{rgb}{1.00,0.45,0.34}
\definecolor{coral2}{rgb}{0.93,0.42,0.31}
\definecolor{coral3}{rgb}{0.80,0.36,0.27}
\definecolor{coral4}{rgb}{0.55,0.24,0.18}
\definecolor{coral}{rgb}{1.00,0.50,0.31}
\definecolor{cornflowerblue}{rgb}{0.39,0.58,0.93}
\definecolor{cornsilk1}{rgb}{1.00,0.97,0.86}
\definecolor{cornsilk2}{rgb}{0.93,0.91,0.80}
\definecolor{cornsilk3}{rgb}{0.80,0.78,0.69}
\definecolor{cornsilk4}{rgb}{0.55,0.53,0.47}
\definecolor{cornsilk}{rgb}{1.00,0.97,0.86}
\definecolor{cyan1}{rgb}{0.00,1.00,1.00}
\definecolor{cyan2}{rgb}{0.00,0.93,0.93}
\definecolor{cyan3}{rgb}{0.00,0.80,0.80}
\definecolor{cyan4}{rgb}{0.00,0.55,0.55}
\definecolor{cyan}{rgb}{0.00,1.00,1.00}
\definecolor{darkblue}{rgb}{0.00,0.00,0.55}
\definecolor{darkcyan}{rgb}{0.00,0.55,0.55}
\definecolor{darkgoldenrod}{rgb}{0.72,0.53,0.04}
\definecolor{darkgray}{rgb}{0.66,0.66,0.66}
\definecolor{darkgreen}{rgb}{0.00,0.39,0.00}
\definecolor{darkgrey}{rgb}{0.66,0.66,0.66}
\definecolor{darkkhaki}{rgb}{0.74,0.72,0.42}
\definecolor{darkmagenta}{rgb}{0.55,0.00,0.55}
\definecolor{darkolive}{rgb}{0.33,0.42,0.18}
\definecolor{darkorange}{rgb}{1.00,0.55,0.00}
\definecolor{darkorchid}{rgb}{0.60,0.20,0.80}
\definecolor{darkred}{rgb}{0.55,0.00,0.00}
\definecolor{darksalmon}{rgb}{0.91,0.59,0.48}
\definecolor{darksea}{rgb}{0.56,0.74,0.56}
\definecolor{darkslate}{rgb}{0.18,0.31,0.31}
\definecolor{darkslate}{rgb}{0.18,0.31,0.31}
\definecolor{darkslate}{rgb}{0.28,0.24,0.55}
\definecolor{darkturquoise}{rgb}{0.00,0.81,0.82}
\definecolor{darkviolet}{rgb}{0.58,0.00,0.83}
\definecolor{deeppink}{rgb}{1.00,0.08,0.58}
\definecolor{deepsky}{rgb}{0.00,0.75,1.00}
\definecolor{dimgray}{rgb}{0.41,0.41,0.41}
\definecolor{dimgrey}{rgb}{0.41,0.41,0.41}
\definecolor{dodgerblue}{rgb}{0.12,0.56,1.00}
\definecolor{firebrick1}{rgb}{1.00,0.19,0.19}
\definecolor{firebrick2}{rgb}{0.93,0.17,0.17}
\definecolor{firebrick3}{rgb}{0.80,0.15,0.15}
\definecolor{firebrick4}{rgb}{0.55,0.10,0.10}
\definecolor{firebrick}{rgb}{0.70,0.13,0.13}
\definecolor{floralwhite}{rgb}{1.00,0.98,0.94}
\definecolor{forestgreen}{rgb}{0.13,0.55,0.13}
\definecolor{gainsboro}{rgb}{0.86,0.86,0.86}
\definecolor{ghostwhite}{rgb}{0.97,0.97,1.00}
\definecolor{gold1}{rgb}{1.00,0.84,0.00}
\definecolor{gold2}{rgb}{0.93,0.79,0.00}
\definecolor{gold3}{rgb}{0.80,0.68,0.00}
\definecolor{gold4}{rgb}{0.55,0.46,0.00}
\definecolor{goldenrod1}{rgb}{1.00,0.76,0.15}
\definecolor{goldenrod2}{rgb}{0.93,0.71,0.13}
\definecolor{goldenrod3}{rgb}{0.80,0.61,0.11}
\definecolor{goldenrod4}{rgb}{0.55,0.41,0.08}
\definecolor{goldenrod}{rgb}{0.85,0.65,0.13}
\definecolor{gold}{rgb}{1.00,0.84,0.00}
\definecolor{gray0}{rgb}{0.00,0.00,0.00}
\definecolor{gray100}{rgb}{1.00,1.00,1.00}
\definecolor{gray10}{rgb}{0.10,0.10,0.10}
\definecolor{gray11}{rgb}{0.11,0.11,0.11}
\definecolor{gray12}{rgb}{0.12,0.12,0.12}
\definecolor{gray13}{rgb}{0.13,0.13,0.13}
\definecolor{gray14}{rgb}{0.14,0.14,0.14}
\definecolor{gray15}{rgb}{0.15,0.15,0.15}
\definecolor{gray16}{rgb}{0.16,0.16,0.16}
\definecolor{gray17}{rgb}{0.17,0.17,0.17}
\definecolor{gray18}{rgb}{0.18,0.18,0.18}
\definecolor{gray19}{rgb}{0.19,0.19,0.19}
\definecolor{gray1}{rgb}{0.01,0.01,0.01}
\definecolor{gray20}{rgb}{0.20,0.20,0.20}
\definecolor{gray21}{rgb}{0.21,0.21,0.21}
\definecolor{gray22}{rgb}{0.22,0.22,0.22}
\definecolor{gray23}{rgb}{0.23,0.23,0.23}
\definecolor{gray24}{rgb}{0.24,0.24,0.24}
\definecolor{gray25}{rgb}{0.25,0.25,0.25}
\definecolor{gray26}{rgb}{0.26,0.26,0.26}
\definecolor{gray27}{rgb}{0.27,0.27,0.27}
\definecolor{gray28}{rgb}{0.28,0.28,0.28}
\definecolor{gray29}{rgb}{0.29,0.29,0.29}
\definecolor{gray2}{rgb}{0.02,0.02,0.02}
\definecolor{gray30}{rgb}{0.30,0.30,0.30}
\definecolor{gray31}{rgb}{0.31,0.31,0.31}
\definecolor{gray32}{rgb}{0.32,0.32,0.32}
\definecolor{gray33}{rgb}{0.33,0.33,0.33}
\definecolor{gray34}{rgb}{0.34,0.34,0.34}
\definecolor{gray35}{rgb}{0.35,0.35,0.35}
\definecolor{gray36}{rgb}{0.36,0.36,0.36}
\definecolor{gray37}{rgb}{0.37,0.37,0.37}
\definecolor{gray38}{rgb}{0.38,0.38,0.38}
\definecolor{gray39}{rgb}{0.39,0.39,0.39}
\definecolor{gray3}{rgb}{0.03,0.03,0.03}
\definecolor{gray40}{rgb}{0.40,0.40,0.40}
\definecolor{gray41}{rgb}{0.41,0.41,0.41}
\definecolor{gray42}{rgb}{0.42,0.42,0.42}
\definecolor{gray43}{rgb}{0.43,0.43,0.43}
\definecolor{gray44}{rgb}{0.44,0.44,0.44}
\definecolor{gray45}{rgb}{0.45,0.45,0.45}
\definecolor{gray46}{rgb}{0.46,0.46,0.46}
\definecolor{gray47}{rgb}{0.47,0.47,0.47}
\definecolor{gray48}{rgb}{0.48,0.48,0.48}
\definecolor{gray49}{rgb}{0.49,0.49,0.49}
\definecolor{gray4}{rgb}{0.04,0.04,0.04}
\definecolor{gray50}{rgb}{0.50,0.50,0.50}
\definecolor{gray51}{rgb}{0.51,0.51,0.51}
\definecolor{gray52}{rgb}{0.52,0.52,0.52}
\definecolor{gray53}{rgb}{0.53,0.53,0.53}
\definecolor{gray54}{rgb}{0.54,0.54,0.54}
\definecolor{gray55}{rgb}{0.55,0.55,0.55}
\definecolor{gray56}{rgb}{0.56,0.56,0.56}
\definecolor{gray57}{rgb}{0.57,0.57,0.57}
\definecolor{gray58}{rgb}{0.58,0.58,0.58}
\definecolor{gray59}{rgb}{0.59,0.59,0.59}
\definecolor{gray5}{rgb}{0.05,0.05,0.05}
\definecolor{gray60}{rgb}{0.60,0.60,0.60}
\definecolor{gray61}{rgb}{0.61,0.61,0.61}
\definecolor{gray62}{rgb}{0.62,0.62,0.62}
\definecolor{gray63}{rgb}{0.63,0.63,0.63}
\definecolor{gray64}{rgb}{0.64,0.64,0.64}
\definecolor{gray65}{rgb}{0.65,0.65,0.65}
\definecolor{gray66}{rgb}{0.66,0.66,0.66}
\definecolor{gray67}{rgb}{0.67,0.67,0.67}
\definecolor{gray68}{rgb}{0.68,0.68,0.68}
\definecolor{gray69}{rgb}{0.69,0.69,0.69}
\definecolor{gray6}{rgb}{0.06,0.06,0.06}
\definecolor{gray70}{rgb}{0.70,0.70,0.70}
\definecolor{gray71}{rgb}{0.71,0.71,0.71}
\definecolor{gray72}{rgb}{0.72,0.72,0.72}
\definecolor{gray73}{rgb}{0.73,0.73,0.73}
\definecolor{gray74}{rgb}{0.74,0.74,0.74}
\definecolor{gray75}{rgb}{0.75,0.75,0.75}
\definecolor{gray76}{rgb}{0.76,0.76,0.76}
\definecolor{gray77}{rgb}{0.77,0.77,0.77}
\definecolor{gray78}{rgb}{0.78,0.78,0.78}
\definecolor{gray79}{rgb}{0.79,0.79,0.79}
\definecolor{gray7}{rgb}{0.07,0.07,0.07}
\definecolor{gray80}{rgb}{0.80,0.80,0.80}
\definecolor{gray81}{rgb}{0.81,0.81,0.81}
\definecolor{gray82}{rgb}{0.82,0.82,0.82}
\definecolor{gray83}{rgb}{0.83,0.83,0.83}
\definecolor{gray84}{rgb}{0.84,0.84,0.84}
\definecolor{gray85}{rgb}{0.85,0.85,0.85}
\definecolor{gray86}{rgb}{0.86,0.86,0.86}
\definecolor{gray87}{rgb}{0.87,0.87,0.87}
\definecolor{gray88}{rgb}{0.88,0.88,0.88}
\definecolor{gray89}{rgb}{0.89,0.89,0.89}
\definecolor{gray8}{rgb}{0.08,0.08,0.08}
\definecolor{gray90}{rgb}{0.90,0.90,0.90}
\definecolor{gray91}{rgb}{0.91,0.91,0.91}
\definecolor{gray92}{rgb}{0.92,0.92,0.92}
\definecolor{gray93}{rgb}{0.93,0.93,0.93}
\definecolor{gray94}{rgb}{0.94,0.94,0.94}
\definecolor{gray95}{rgb}{0.95,0.95,0.95}
\definecolor{gray96}{rgb}{0.96,0.96,0.96}
\definecolor{gray97}{rgb}{0.97,0.97,0.97}
\definecolor{gray98}{rgb}{0.98,0.98,0.98}
\definecolor{gray99}{rgb}{0.99,0.99,0.99}
\definecolor{gray9}{rgb}{0.09,0.09,0.09}
\definecolor{gray}{rgb}{0.75,0.75,0.75}
\definecolor{green1}{rgb}{0.00,1.00,0.00}
\definecolor{green2}{rgb}{0.00,0.93,0.00}
\definecolor{green3}{rgb}{0.00,0.80,0.00}
\definecolor{green4}{rgb}{0.00,0.55,0.00}
\definecolor{greenyellow}{rgb}{0.68,1.00,0.18}
\definecolor{green}{rgb}{0.00,1.00,0.00}
\definecolor{grey0}{rgb}{0.00,0.00,0.00}
\definecolor{grey100}{rgb}{1.00,1.00,1.00}
\definecolor{grey10}{rgb}{0.10,0.10,0.10}
\definecolor{grey11}{rgb}{0.11,0.11,0.11}
\definecolor{grey12}{rgb}{0.12,0.12,0.12}
\definecolor{grey13}{rgb}{0.13,0.13,0.13}
\definecolor{grey14}{rgb}{0.14,0.14,0.14}
\definecolor{grey15}{rgb}{0.15,0.15,0.15}
\definecolor{grey16}{rgb}{0.16,0.16,0.16}
\definecolor{grey17}{rgb}{0.17,0.17,0.17}
\definecolor{grey18}{rgb}{0.18,0.18,0.18}
\definecolor{grey19}{rgb}{0.19,0.19,0.19}
\definecolor{grey1}{rgb}{0.01,0.01,0.01}
\definecolor{grey20}{rgb}{0.20,0.20,0.20}
\definecolor{grey21}{rgb}{0.21,0.21,0.21}
\definecolor{grey22}{rgb}{0.22,0.22,0.22}
\definecolor{grey23}{rgb}{0.23,0.23,0.23}
\definecolor{grey24}{rgb}{0.24,0.24,0.24}
\definecolor{grey25}{rgb}{0.25,0.25,0.25}
\definecolor{grey26}{rgb}{0.26,0.26,0.26}
\definecolor{grey27}{rgb}{0.27,0.27,0.27}
\definecolor{grey28}{rgb}{0.28,0.28,0.28}
\definecolor{grey29}{rgb}{0.29,0.29,0.29}
\definecolor{grey2}{rgb}{0.02,0.02,0.02}
\definecolor{grey30}{rgb}{0.30,0.30,0.30}
\definecolor{grey31}{rgb}{0.31,0.31,0.31}
\definecolor{grey32}{rgb}{0.32,0.32,0.32}
\definecolor{grey33}{rgb}{0.33,0.33,0.33}
\definecolor{grey34}{rgb}{0.34,0.34,0.34}
\definecolor{grey35}{rgb}{0.35,0.35,0.35}
\definecolor{grey36}{rgb}{0.36,0.36,0.36}
\definecolor{grey37}{rgb}{0.37,0.37,0.37}
\definecolor{grey38}{rgb}{0.38,0.38,0.38}
\definecolor{grey39}{rgb}{0.39,0.39,0.39}
\definecolor{grey3}{rgb}{0.03,0.03,0.03}
\definecolor{grey40}{rgb}{0.40,0.40,0.40}
\definecolor{grey41}{rgb}{0.41,0.41,0.41}
\definecolor{grey42}{rgb}{0.42,0.42,0.42}
\definecolor{grey43}{rgb}{0.43,0.43,0.43}
\definecolor{grey44}{rgb}{0.44,0.44,0.44}
\definecolor{grey45}{rgb}{0.45,0.45,0.45}
\definecolor{grey46}{rgb}{0.46,0.46,0.46}
\definecolor{grey47}{rgb}{0.47,0.47,0.47}
\definecolor{grey48}{rgb}{0.48,0.48,0.48}
\definecolor{grey49}{rgb}{0.49,0.49,0.49}
\definecolor{grey4}{rgb}{0.04,0.04,0.04}
\definecolor{grey50}{rgb}{0.50,0.50,0.50}
\definecolor{grey51}{rgb}{0.51,0.51,0.51}
\definecolor{grey52}{rgb}{0.52,0.52,0.52}
\definecolor{grey53}{rgb}{0.53,0.53,0.53}
\definecolor{grey54}{rgb}{0.54,0.54,0.54}
\definecolor{grey55}{rgb}{0.55,0.55,0.55}
\definecolor{grey56}{rgb}{0.56,0.56,0.56}
\definecolor{grey57}{rgb}{0.57,0.57,0.57}
\definecolor{grey58}{rgb}{0.58,0.58,0.58}
\definecolor{grey59}{rgb}{0.59,0.59,0.59}
\definecolor{grey5}{rgb}{0.05,0.05,0.05}
\definecolor{grey60}{rgb}{0.60,0.60,0.60}
\definecolor{grey61}{rgb}{0.61,0.61,0.61}
\definecolor{grey62}{rgb}{0.62,0.62,0.62}
\definecolor{grey63}{rgb}{0.63,0.63,0.63}
\definecolor{grey64}{rgb}{0.64,0.64,0.64}
\definecolor{grey65}{rgb}{0.65,0.65,0.65}
\definecolor{grey66}{rgb}{0.66,0.66,0.66}
\definecolor{grey67}{rgb}{0.67,0.67,0.67}
\definecolor{grey68}{rgb}{0.68,0.68,0.68}
\definecolor{grey69}{rgb}{0.69,0.69,0.69}
\definecolor{grey6}{rgb}{0.06,0.06,0.06}
\definecolor{grey70}{rgb}{0.70,0.70,0.70}
\definecolor{grey71}{rgb}{0.71,0.71,0.71}
\definecolor{grey72}{rgb}{0.72,0.72,0.72}
\definecolor{grey73}{rgb}{0.73,0.73,0.73}
\definecolor{grey74}{rgb}{0.74,0.74,0.74}
\definecolor{grey75}{rgb}{0.75,0.75,0.75}
\definecolor{grey76}{rgb}{0.76,0.76,0.76}
\definecolor{grey77}{rgb}{0.77,0.77,0.77}
\definecolor{grey78}{rgb}{0.78,0.78,0.78}
\definecolor{grey79}{rgb}{0.79,0.79,0.79}
\definecolor{grey7}{rgb}{0.07,0.07,0.07}
\definecolor{grey80}{rgb}{0.80,0.80,0.80}
\definecolor{grey81}{rgb}{0.81,0.81,0.81}
\definecolor{grey82}{rgb}{0.82,0.82,0.82}
\definecolor{grey83}{rgb}{0.83,0.83,0.83}
\definecolor{grey84}{rgb}{0.84,0.84,0.84}
\definecolor{grey85}{rgb}{0.85,0.85,0.85}
\definecolor{grey86}{rgb}{0.86,0.86,0.86}
\definecolor{grey87}{rgb}{0.87,0.87,0.87}
\definecolor{grey88}{rgb}{0.88,0.88,0.88}
\definecolor{grey89}{rgb}{0.89,0.89,0.89}
\definecolor{grey8}{rgb}{0.08,0.08,0.08}
\definecolor{grey90}{rgb}{0.90,0.90,0.90}
\definecolor{grey91}{rgb}{0.91,0.91,0.91}
\definecolor{grey92}{rgb}{0.92,0.92,0.92}
\definecolor{grey93}{rgb}{0.93,0.93,0.93}
\definecolor{grey94}{rgb}{0.94,0.94,0.94}
\definecolor{grey95}{rgb}{0.95,0.95,0.95}
\definecolor{grey96}{rgb}{0.96,0.96,0.96}
\definecolor{grey97}{rgb}{0.97,0.97,0.97}
\definecolor{grey98}{rgb}{0.98,0.98,0.98}
\definecolor{grey99}{rgb}{0.99,0.99,0.99}
\definecolor{grey9}{rgb}{0.09,0.09,0.09}
\definecolor{grey}{rgb}{0.75,0.75,0.75}
\definecolor{honeydew1}{rgb}{0.94,1.00,0.94}
\definecolor{honeydew2}{rgb}{0.88,0.93,0.88}
\definecolor{honeydew3}{rgb}{0.76,0.80,0.76}
\definecolor{honeydew4}{rgb}{0.51,0.55,0.51}
\definecolor{honeydew}{rgb}{0.94,1.00,0.94}
\definecolor{hotpink}{rgb}{1.00,0.41,0.71}
\definecolor{indianred}{rgb}{0.80,0.36,0.36}
\definecolor{ivory1}{rgb}{1.00,1.00,0.94}
\definecolor{ivory2}{rgb}{0.93,0.93,0.88}
\definecolor{ivory3}{rgb}{0.80,0.80,0.76}
\definecolor{ivory4}{rgb}{0.55,0.55,0.51}
\definecolor{ivory}{rgb}{1.00,1.00,0.94}
\definecolor{khaki1}{rgb}{1.00,0.96,0.56}
\definecolor{khaki2}{rgb}{0.93,0.90,0.52}
\definecolor{khaki3}{rgb}{0.80,0.78,0.45}
\definecolor{khaki4}{rgb}{0.55,0.53,0.31}
\definecolor{khaki}{rgb}{0.94,0.90,0.55}
\definecolor{lavenderblush}{rgb}{1.00,0.94,0.96}
\definecolor{lavender}{rgb}{0.90,0.90,0.98}
\definecolor{lawngreen}{rgb}{0.49,0.99,0.00}
\definecolor{lemonchiffon}{rgb}{1.00,0.98,0.80}
\definecolor{lightblue}{rgb}{0.68,0.85,0.90}
\definecolor{lightcoral}{rgb}{0.94,0.50,0.50}
\definecolor{lightcyan}{rgb}{0.88,1.00,1.00}
\definecolor{lightgoldenrod}{rgb}{0.93,0.87,0.51}
\definecolor{lightgoldenrod}{rgb}{0.98,0.98,0.82}
\definecolor{lightgray}{rgb}{0.83,0.83,0.83}
\definecolor{lightgreen}{rgb}{0.56,0.93,0.56}
\definecolor{lightgrey}{rgb}{0.83,0.83,0.83}
\definecolor{lightpink}{rgb}{1.00,0.71,0.76}
\definecolor{lightsalmon}{rgb}{1.00,0.63,0.48}
\definecolor{lightsea}{rgb}{0.13,0.70,0.67}
\definecolor{lightsky}{rgb}{0.53,0.81,0.98}
\definecolor{lightslate}{rgb}{0.47,0.53,0.60}
\definecolor{lightslate}{rgb}{0.47,0.53,0.60}
\definecolor{lightslate}{rgb}{0.52,0.44,1.00}
\definecolor{lightsteel}{rgb}{0.69,0.77,0.87}
\definecolor{lightyellow}{rgb}{1.00,1.00,0.88}
\definecolor{limegreen}{rgb}{0.20,0.80,0.20}
\definecolor{linen}{rgb}{0.98,0.94,0.90}
\definecolor{magenta1}{rgb}{1.00,0.00,1.00}
\definecolor{magenta2}{rgb}{0.93,0.00,0.93}
\definecolor{magenta3}{rgb}{0.80,0.00,0.80}
\definecolor{magenta4}{rgb}{0.55,0.00,0.55}
\definecolor{magenta}{rgb}{1.00,0.00,1.00}
\definecolor{maroon1}{rgb}{1.00,0.20,0.70}
\definecolor{maroon2}{rgb}{0.93,0.19,0.65}
\definecolor{maroon3}{rgb}{0.80,0.16,0.56}
\definecolor{maroon4}{rgb}{0.55,0.11,0.38}
\definecolor{maroon}{rgb}{0.69,0.19,0.38}
\definecolor{mediumaquamarine}{rgb}{0.40,0.80,0.67}
\definecolor{mediumblue}{rgb}{0.00,0.00,0.80}
\definecolor{mediumorchid}{rgb}{0.73,0.33,0.83}
\definecolor{mediumpurple}{rgb}{0.58,0.44,0.86}
\definecolor{mediumsea}{rgb}{0.24,0.70,0.44}
\definecolor{mediumslate}{rgb}{0.48,0.41,0.93}
\definecolor{mediumspring}{rgb}{0.00,0.98,0.60}
\definecolor{mediumturquoise}{rgb}{0.28,0.82,0.80}
\definecolor{mediumviolet}{rgb}{0.78,0.08,0.52}
\definecolor{midnightblue}{rgb}{0.10,0.10,0.44}
\definecolor{mintcream}{rgb}{0.96,1.00,0.98}
\definecolor{mistyrose}{rgb}{1.00,0.89,0.88}
\definecolor{moccasin}{rgb}{1.00,0.89,0.71}
\definecolor{navajowhite}{rgb}{1.00,0.87,0.68}
\definecolor{navyblue}{rgb}{0.00,0.00,0.50}
\definecolor{navy}{rgb}{0.00,0.00,0.50}
\definecolor{oldlace}{rgb}{0.99,0.96,0.90}
\definecolor{olivedrab}{rgb}{0.42,0.56,0.14}
\definecolor{orange1}{rgb}{1.00,0.65,0.00}
\definecolor{orange2}{rgb}{0.93,0.60,0.00}
\definecolor{orange3}{rgb}{0.80,0.52,0.00}
\definecolor{orange4}{rgb}{0.55,0.35,0.00}
\definecolor{orangered}{rgb}{1.00,0.27,0.00}
\definecolor{orange}{rgb}{1.00,0.65,0.00}
\definecolor{orchid1}{rgb}{1.00,0.51,0.98}
\definecolor{orchid2}{rgb}{0.93,0.48,0.91}
\definecolor{orchid3}{rgb}{0.80,0.41,0.79}
\definecolor{orchid4}{rgb}{0.55,0.28,0.54}
\definecolor{orchid}{rgb}{0.85,0.44,0.84}
\definecolor{palegoldenrod}{rgb}{0.93,0.91,0.67}
\definecolor{palegreen}{rgb}{0.60,0.98,0.60}
\definecolor{paleturquoise}{rgb}{0.69,0.93,0.93}
\definecolor{paleviolet}{rgb}{0.86,0.44,0.58}
\definecolor{papayawhip}{rgb}{1.00,0.94,0.84}
\definecolor{peachpuff}{rgb}{1.00,0.85,0.73}
\definecolor{peru}{rgb}{0.80,0.52,0.25}
\definecolor{pink1}{rgb}{1.00,0.71,0.77}
\definecolor{pink2}{rgb}{0.93,0.66,0.72}
\definecolor{pink3}{rgb}{0.80,0.57,0.62}
\definecolor{pink4}{rgb}{0.55,0.39,0.42}
\definecolor{pink}{rgb}{1.00,0.75,0.80}
\definecolor{plum1}{rgb}{1.00,0.73,1.00}
\definecolor{plum2}{rgb}{0.93,0.68,0.93}
\definecolor{plum3}{rgb}{0.80,0.59,0.80}
\definecolor{plum4}{rgb}{0.55,0.40,0.55}
\definecolor{plum}{rgb}{0.87,0.63,0.87}
\definecolor{powderblue}{rgb}{0.69,0.88,0.90}
\definecolor{purple1}{rgb}{0.61,0.19,1.00}
\definecolor{purple2}{rgb}{0.57,0.17,0.93}
\definecolor{purple3}{rgb}{0.49,0.15,0.80}
\definecolor{purple4}{rgb}{0.33,0.10,0.55}
\definecolor{purple}{rgb}{0.63,0.13,0.94}
\definecolor{red1}{rgb}{1.00,0.00,0.00}
\definecolor{red2}{rgb}{0.93,0.00,0.00}
\definecolor{red3}{rgb}{0.80,0.00,0.00}
\definecolor{red4}{rgb}{0.55,0.00,0.00}
\definecolor{red}{rgb}{1.00,0.00,0.00}
\definecolor{rosybrown}{rgb}{0.74,0.56,0.56}
\definecolor{royalblue}{rgb}{0.25,0.41,0.88}
\definecolor{saddlebrown}{rgb}{0.55,0.27,0.07}
\definecolor{salmon1}{rgb}{1.00,0.55,0.41}
\definecolor{salmon2}{rgb}{0.93,0.51,0.38}
\definecolor{salmon3}{rgb}{0.80,0.44,0.33}
\definecolor{salmon4}{rgb}{0.55,0.30,0.22}
\definecolor{salmon}{rgb}{0.98,0.50,0.45}
\definecolor{sandybrown}{rgb}{0.96,0.64,0.38}
\definecolor{seagreen}{rgb}{0.18,0.55,0.34}
\definecolor{seashell1}{rgb}{1.00,0.96,0.93}
\definecolor{seashell2}{rgb}{0.93,0.90,0.87}
\definecolor{seashell3}{rgb}{0.80,0.77,0.75}
\definecolor{seashell4}{rgb}{0.55,0.53,0.51}
\definecolor{seashell}{rgb}{1.00,0.96,0.93}
\definecolor{sienna1}{rgb}{1.00,0.51,0.28}
\definecolor{sienna2}{rgb}{0.93,0.47,0.26}
\definecolor{sienna3}{rgb}{0.80,0.41,0.22}
\definecolor{sienna4}{rgb}{0.55,0.28,0.15}
\definecolor{sienna}{rgb}{0.63,0.32,0.18}
\definecolor{skyblue}{rgb}{0.53,0.81,0.92}
\definecolor{slateblue}{rgb}{0.42,0.35,0.80}
\definecolor{slategray}{rgb}{0.44,0.50,0.56}
\definecolor{slategrey}{rgb}{0.44,0.50,0.56}
\definecolor{snow1}{rgb}{1.00,0.98,0.98}
\definecolor{snow2}{rgb}{0.93,0.91,0.91}
\definecolor{snow3}{rgb}{0.80,0.79,0.79}
\definecolor{snow4}{rgb}{0.55,0.54,0.54}
\definecolor{snow}{rgb}{1.00,0.98,0.98}
\definecolor{springgreen}{rgb}{0.00,1.00,0.50}
\definecolor{steelblue}{rgb}{0.27,0.51,0.71}
\definecolor{tan1}{rgb}{1.00,0.65,0.31}
\definecolor{tan2}{rgb}{0.93,0.60,0.29}
\definecolor{tan3}{rgb}{0.80,0.52,0.25}
\definecolor{tan4}{rgb}{0.55,0.35,0.17}
\definecolor{tan}{rgb}{0.82,0.71,0.55}
\definecolor{thistle1}{rgb}{1.00,0.88,1.00}
\definecolor{thistle2}{rgb}{0.93,0.82,0.93}
\definecolor{thistle3}{rgb}{0.80,0.71,0.80}
\definecolor{thistle4}{rgb}{0.55,0.48,0.55}
\definecolor{thistle}{rgb}{0.85,0.75,0.85}
\definecolor{tomato1}{rgb}{1.00,0.39,0.28}
\definecolor{tomato2}{rgb}{0.93,0.36,0.26}
\definecolor{tomato3}{rgb}{0.80,0.31,0.22}
\definecolor{tomato4}{rgb}{0.55,0.21,0.15}
\definecolor{tomato}{rgb}{1.00,0.39,0.28}
\definecolor{turquoise1}{rgb}{0.00,0.96,1.00}
\definecolor{turquoise2}{rgb}{0.00,0.90,0.93}
\definecolor{turquoise3}{rgb}{0.00,0.77,0.80}
\definecolor{turquoise4}{rgb}{0.00,0.53,0.55}
\definecolor{turquoise}{rgb}{0.25,0.88,0.82}
\definecolor{violetred}{rgb}{0.82,0.13,0.56}
\definecolor{violet}{rgb}{0.93,0.51,0.93}
\definecolor{wheat1}{rgb}{1.00,0.91,0.73}
\definecolor{wheat2}{rgb}{0.93,0.85,0.68}
\definecolor{wheat3}{rgb}{0.80,0.73,0.59}
\definecolor{wheat4}{rgb}{0.55,0.49,0.40}
\definecolor{wheat}{rgb}{0.96,0.87,0.70}
\definecolor{whitesmoke}{rgb}{0.96,0.96,0.96}
\definecolor{white}{rgb}{1.00,1.00,1.00}
\definecolor{yellow1}{rgb}{1.00,1.00,0.00}
\definecolor{yellow2}{rgb}{0.93,0.93,0.00}
\definecolor{yellow3}{rgb}{0.80,0.80,0.00}
\definecolor{yellow4}{rgb}{0.55,0.55,0.00}
\definecolor{yellowgreen}{rgb}{0.60,0.80,0.20}
\definecolor{yellow}{rgb}{1.00,1.00,0.00}

%% file: introduction.tex
\section{Introduction} 
Much conceptual, analytical, and empirical research has been conducted during the
last decades to enhance our understanding of conceptual modeling. In particular,
process models have gained significant importance due to their fundamental role
for process-aware information systems \cite{Bec+00}. Even though it is well known
that a good understanding of a process model has a direct and measurable impact
on the success of any modeling initiative \cite{DBLP:journals/dss/KockVDD09},
process models display a wide range of quality problems impeding their
comprehensibility and hampering their maintainability
\cite{DBLP:journals/dke/MendlingVDAN08,WRR08}. Literature reports, for example,
on error rates between 10\% and 20\% in industrial process model collections
\cite{DBLP:journals/dke/MendlingVDAN08}.

To improve process model quality, change patterns offer a promising
perspective~\cite{WRR08}. They have well-defined semantics
\cite{DBLP:conf/er/Rinderle-MaRW08} and combine change primitives (e.g., to add
nodes or edges) to high-level change operations. Particularly appealing is
correctness-by-construction \cite{ReDa98,Casa98}, i.e., the modeling environment
only provides those change patterns to the process designers, which ensure
that a sound process model is transformed into another sound process model.


The use of change patterns implies a different way of creating process models,
since correctness-by-construction imposes a structured way of modeling by
enforcing block structuredness. Irrespective of whether change patterns or change
primitives are used, model creation requires process designers to construct a
mental model (i.e., \textit{internal representation}) of the requirements to be
captured in the process model~\cite{Sof+12}. In a subsequent step, the mental
model is mapped to the constructs provided by the modeling language creating an
\textit{external representation} of the domain~\cite{Sof+12}. While the
construction of the mental model presumably remains unaffected, the use of change
patterns leads to different challenges concerning pattern selection and
combination for creating the external representation. Further, the exact set of
change patterns available influences the selection.  While a large set of change
patterns allows for more flexibility, it also increases complexity, making the
modeling environment more difficult to use. Consequently, process designers might
have to look several steps ahead to construct a certain process fragment, which
constitutes a major difference compared to the use of change primitives, which do
not impose any structural restrictions (i.e., no order is imposed when placing
elements on the modeling canvas).

The process of creating process models based on change primitives has caused
significant attention leading to a stream of research on the \textit{process of
process modeling} (PPM) \cite{Sof+12,PZW+12,CVR+12,PSZ+12}. This research is
characterized by its focus on the formalization phase of model creation, i.e.,
the designer's interactions with the modeling environment~\cite{PZW+12}. Still,
little is known about the PPM when utilizing change patterns. In this paper, we
try to obtain an in-depth understanding of the PPM when using change patterns. In
particular, this paper aims at understanding whether the necessity to look ahead
leads to additional barriers during model creation and at shedding light on
the challenges process designers face when using change patterns, which is
essential to provide effective tool support.

To obtain an in-depth understanding of these issues, we implement a change
pattern modeling editor based on Cheetah Experimental Platform (CEP) and
conduct an exploratory study comprising several modeling tasks. The results of
our study underline the potential for model creation based on change patterns. In
particular, process designers did not face any major problem when using change
patterns for constructing simple process fragments. When more complex process
fragments forced process designers to look ahead, difficulties increased
observably. Insights obtained from this exploratory study provide a better understanding of
the PPM when using changes patterns and reveal challenges arising in such a
context. In particular, results provide a contribution toward a better
understanding on how tool features like change patterns impact the PPM, but also
give advice on how effective tool support should be designed.

Sect. \ref{backgrounds} introduces backgrounds. Sect. \ref{exploratoryStudy} 
describes the design of the exploratory study. Sect. \ref{challenges} discusses
challenges in change pattern use and Sect. \ref{easeOfUse} presents results
on process designers' perception of patterns use. Results are
discussed in Sect. \ref{discuss}. Related work is presented in Sect.
\ref{relatedWork}. Sect. \ref{summary} concludes the paper.

%% file: backgrounds-vt.tex
\section{Backgrounds}
\label{backgrounds}
This section briefly introduces change patterns and then presents cognitive
foundations and backgrounds of the PPM.
 
\subsection{Process Modeling Based on Change Patterns}
\label{patternsVersusPrimitives}
\begin{figure}[h!tb]
\vspace{-4mm}    
\center
  \includegraphics[width=1\textwidth]{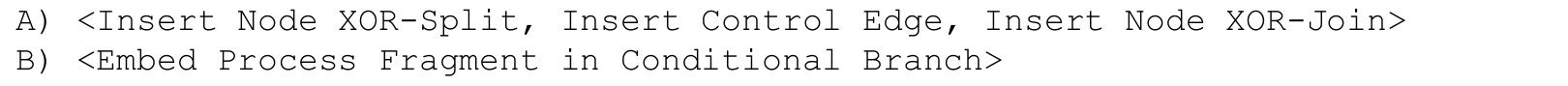} \caption{Set
  of Change Primitives (A) and Patterns (B) to make an activity optional}
  \label{img:primitivesVsPatterns}
  \vspace{-6mm}
  \center
\end{figure}

Change patterns provide a different way of interacting with the modeling
environment in modeling phases. Instead of specifying a set of change primitives,
the process designer applies change patterns (i.e., high-level change operations)
to realize the desired model adaptation (cf. Fig.
\ref{img:primitivesVsPatterns}). Examples of change patterns include the
insertion and deletion of process fragments, or their embedding in
loops (the whole catalogue can be found in \cite{DaRe09,WRR08} and their
semantics in \cite{DBLP:conf/er/Rinderle-MaRW08}). When applying a single
change primitive, soundness of the resulting process model cannot be guaranteed, but
must be explicitly checked after applying the change primitives. On the contrary,
change pattern implementations often associate pre-/post-conditions with
high-level change operations to guarantee model correctness after each adaptation
\cite{ReDa98,Casa98}. Process editors applying the correctness-by-construction
principle (e.g., \cite{DaRe09}) usually provide only those change patterns to the
process designer that allow transforming a sound process model into another sound
one. This is realized by imposing structural restrictions on process models
(e.g., block structuredness).
     
\subsection{Cognitive Foundations of Problem Solving}\label{cognFound}
We consider the creation of process models to be a complex problem solving task.
Problem solving has been an area of vivid research in cognitive psychology for
decades. Therefore, we turn to cognitive psychology to understand the processes
followed by humans when solving a problem like creating a process model.
 
\textbf{Schemata.} A central insight from cognitive research is that the human
brain contains specialized regions contributing different functionality to the
process of \emph{solving complex problems}. \emph{Long-term memory} is
responsible for permanently storing information and has an essentially unlimited
capacity, while in \emph{working memory} comparing, computing and reasoning take
place~\cite{Gray07}. Although the latter is the main working area of the brain, it can store only a limited amount of information, which is forgotten after 20--30
seconds if not refreshed~\cite{Trac79}. The question arises how information can
be processed with such limited capacity. The human mind organizes information in
interconnected \emph{schemata} rather than in isolation~\cite{Gray07}. Those
schemata, stored in long-term memory, incorporate general concepts of similar
situations~\cite{Gray07}. Whenever situations similar to a schema arise, the
latter is retrieved to help organizing information by creating \emph{chunks} of
information that can be processed efficiently~\cite{JTPA81}. For instance, when
asked to create a process model using change patterns, a designer must create an
internal representation of the problem~\cite{Sof+12}. For this, information about
the domain is retrieved and organized in memory using existing schemata for
process modeling. Therefore, schemata guide the comprehension process, helping to
re-organize information for its processing in working memory~\cite{JTPA81}.

\textbf{Problem-Solving Strategies.} When
confronting novices with an unfamiliar problem they cannot rely on specialized
problem solving strategies. Instead, they must find a way to solve the problem
and come up with an initial skeletal \emph{plan}~\cite{Rist89}. Then, novices
utilize general problem solving strategies, like means-ends analysis, due to
the lack of more specific strategies for the task at hand~\cite{KaNe84}.
Means-ends analysis can be described as the continual comparison of the problem's current
state with the desired end product. Based on this, the next steps are selected until a
satisfying solution is found~\cite{KaNe84}. After applying the constructed plan,
it can be stored in long-term memory as \emph{plan schema}~\cite{Rist89}. For
this, task-specific details are removed from the plan schema resulting in a plan
schema that can be automatically applied in similar situations~\cite{Ande82}.
When confronted with a problem solving task in the future, the appropriate plan
schema is selected using a case-based reasoning approach~\cite{GuCu88}. The
retrieved plan schema provides the user with structured knowledge
that drives the process of solving the problem~\cite{JTPA81,GuCu88}. Plan
schemata allow experts to immediately decide what steps to apply to end up with
the desired solution~\cite{Swel88}. If the plan schema is well developed, an
expert never reaches a dead end when solving the problem~\cite{Broo77}.

Plan schemata seem to be important when creating process models based on change
patterns since change pattern cannot be combined in an arbitrary manner. If no
plan schema is available on how to combine change patterns to create the desired
process model, designers have to utilize means-ends analysis until a satisfying
solution is found. This behavior is more likely to reach in detours, reducing the
process designer's efficiency when creating process models.

\subsection{The Process of Process Modeling}\label{PPM}
During the formalization phase, process designers create a
syntactically correct process model reflecting a given domain description by interacting with the
process editor~\cite{HoPW05}. The formalization, i.e., the PPM, can be described
as a cycle of the three phases of comprehension, modeling and
reconciliation~\cite{PZW+12,Sof+12}.

\textbf{Comprehension.} In comprehension phases designers try to understand the
requirements to be modeled by extracting information from the task description
and the existing process model to build an internal representation of the problem
in working memory~\cite{KaNe84,Broo77}. Depending on the availability of schemata
for organizing the acquired knowledge, working memory is utilized more or less
efficiently. If the process designer has solved a similar problem previously
(i.e., a plan schema for the problem is stored in long-term memory), he can
directly create the process model without any further attention on which steps
to execute or plan next. Either way, working memory is filled with
knowledge extracted from requirements and, if available, the process model being created.


 \textbf{Modeling.} The designer uses the internal representation
developed during the comprehension phase to materialize the solution in a process
model (by creating or changing it)~\cite{Sof+12,PZW+12}. Hence, modeling phases
consist of a set of structural model adaptations. More specifically, designers
interact with the modeling environment using change primitives such as {\tt add
activity} or {\tt add edge}. The materialization of a process model adaptation
may require the joint application of several change primitives. For instance,
making an activity optional may require the application of three change
primitives as shown in Fig. \ref{img:primitivesVsPatterns}A. Once all information stored in working memory is incorporated in the process model, the
designer interrupts the modeling endeavor to incorporate additional requirements
into the internal representation~\cite{PZW+12}.

\textbf{Reconciliation.} Designers may reorganize a process model (e.g.,\
renaming of activities) and utilize its \emph{secondary notation} (e.g.,\
notation of layout, typographic cues) to enhance understandability~\cite{Petr95}. However, the
number of reconciliation phases in the PPM depends on the designer's ability
of placing elements correctly when creating them~\cite{PZW+12}.
 
\vspace{-0.2cm}

%% file: casestudy.tex
\vspace{-0.2cm}
\section{Exploratory Study}
\label{exploratoryStudy}
To gain a better understanding of the PPM using change
patterns we conduct an exploratory study. This section
describes research questions and study design. 

\textbf{Research Questions.} Central aim is to obtain an in-depth understanding
of the PPM when using change patterns. More specifically, we try to understand the
challenges designers are facing when using change patterns for model creation.
Respective challenges can result in detours for designers on their way to a
complete process model. Detours during process modeling result in decreased
problem solving efficiency. Moreover, they might lead to modeling errors that
persist in the final model.

\vspace{0.1cm}
\begin{center}
\fbox{
  \parbox{11cm}{
 RQ1: What are re-occurring challenges in the usage of change
  patterns that designers face and where do these challenges originate from?
  }
 } 
\end{center}
\vspace{0.1cm}

\noindent The study also aims at investigating how designers experience
their interaction with the modeling environment to create process
models using change pattern. 

\begin{center}
\fbox{
  \parbox{11cm}{
 RQ2: What is the subjective perception of designers when using change
  patterns for model creation?
  }
 } 
\end{center}

\textbf{Exploratory Study Execution.} The design of the study consists
of three phases. In the first phase, demographic data is collected. In the second
phase, two modeling tasks are executed. When
working on the modeling tasks all interactions with the modeling environment 
are recorded using CEP~\cite{PZW10}. This allows us to replay the creation of the process model
step-by-step (cf.~\cite{PZW10,PZW+12}), addressing RQ1. After
completing the modeling tasks, \textit{Perceived Ease of Use} and the
\textit{Perceived Usefulness} of \textit{Technology Acceptance Model
(TAM)}~\cite{Davis89} are assessed to investigate RQ2.
In addition, participating subjects are asked to provide feedback regarding
their experiences.

\textbf{Subjects.} The exploratory study was conducted in Innsbruck with
16 students of a graduate course on business process management. Since designers
in practical settings are often not expert designers, but rather casual designers
that only obtained a basic amount of training~\cite{PZW+10}, we did not require
modeling experts for our study. Subjects obtained basic training in modeling
business processes prior to the exploratory study. In addition, they were
taught theoretical backgrounds on change patterns prior to the exploratory
study, but did not have any hands-on experience in the creation of process models
using change patterns.

\textbf{Modeling Tasks.} 
For Task A, subjects received an informal requirements description and the
solution of the modeling task (i.e., a process model). Subjects had to re-model
the process using change patterns. Since subjects had the correct solution
available, the challenge lies in determining the patterns to use for
re-building the model and how to combine them effectively. This
allowed students to develop problem solving strategies for utilizing
change pattern. Task A was a process run by the ``Task Force Earthquakes" of the
German Research Center for Geosciences~\cite{DBLP:conf/bpm/FahlandW08}. Subjects
were asked to model the ``Transport of Equipment" process using change patterns.
The task requires sequences as well as conditional/parallel branchings. The
solution model has a nesting depth of 2.

In Task B, designers had to create a process model starting from an
informal description. This time, no solution model was made available to the
subjects. Consequently, they not only had to decide which patterns to use
and how to combine them, but additionally had to develop an understanding of the
domain (by creating a mental model) and map it to the available change
patterns. Therefore, schemata for extracting information from the textual description were necessary for completing the modeling task
(e.g., for identifying activities). Task B describes the pre-take off procedures
for a general aviation flight under visual rules~\cite{ReWe12}. Like Task A, it
comprises sequences and conditional/parallel branchings.
In terms of complexity, it only has a nesting depth of 1.

\textbf{Change Pattern Set.} 
When devising a modeling environment for model creation based on change patterns
the question arises which change pattern set to provide. While a large set offers
more flexibility, it also increases complexity---especially when mapping the
mental model to the available patterns. We therefore utilize a minimal change
pattern set (for the full pattern set see \cite{WRR08}), which allows designers
to create all main control-flow constructs for tasks A and B (i.e., sequences,
parallel branches, conditional branchings, and loops). The following patterns were available to the
designers: AP1 (Insert Process Fragment), AP2 (Delete Process Fragment), AP8
(Embed Fragment in Loop), AP10 (Embed Process Fragment in Conditional Branch),
and AP13 (Update Condition). Concerning AP1, two pattern variants were provided:
Serial Insert and Parallel Insert.


%% file: results.tex
\vspace{-5mm} 

\section{Challenges in the Usage of Change Patterns}
\label{challenges}

This section describes the data analysis procedure applied to obtain an in-depth
understanding of re-occurring challenges during the PPM based on change patterns,
answering RQ1 it further presents obtained results for Tasks A and B.

\vspace{-2mm} 
  
\subsection{Data Analysis Procedure}  \label{dataAnalysis}
 
\begin{figure}[h]
\begin{center}
\includegraphics[width=0.8\textwidth]{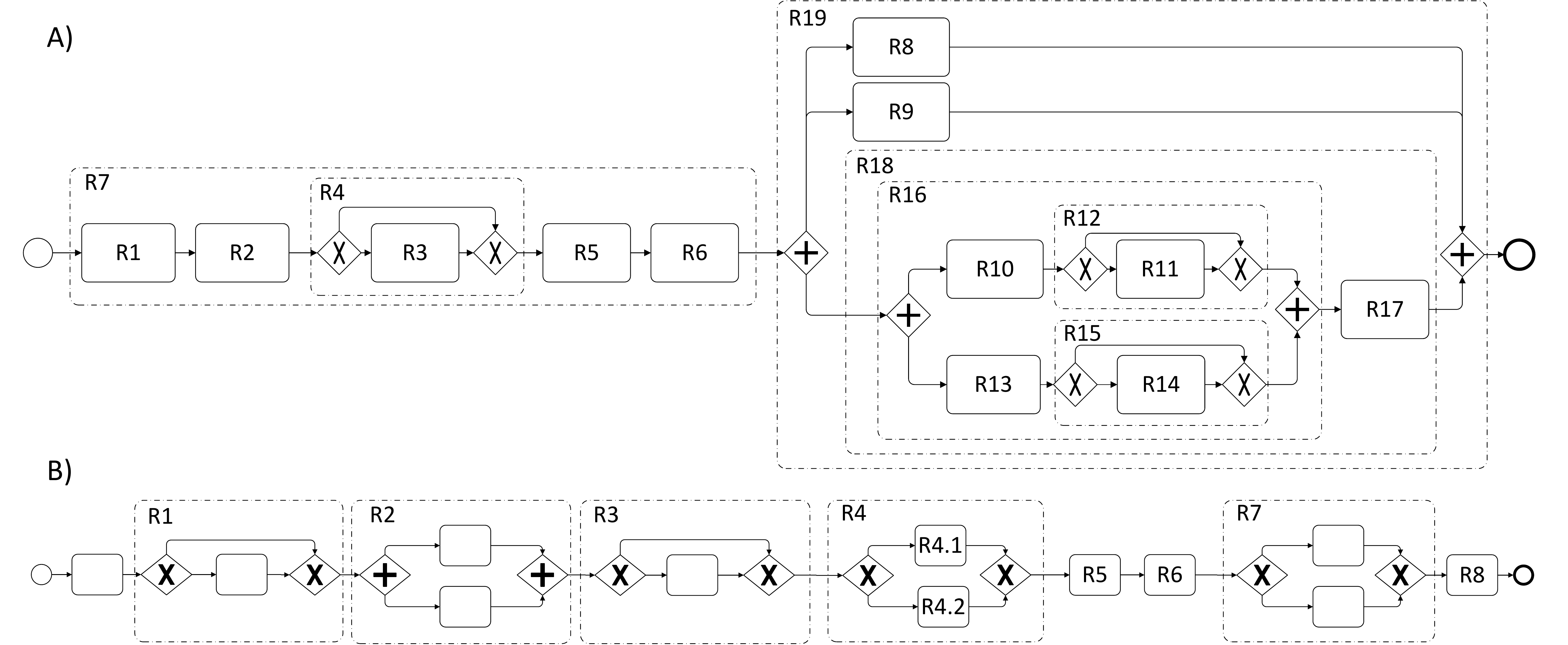}
\caption{Solution Models for Tasks A and B}\label{img:tasksA+B}
\vspace{-6mm}  
\end{center}
\end{figure}

 
\textit{Step 1: Determine Solution Model, Distance, and Optimal Problem Solving
Paths. } In a first step, for each modeling task, we create a model representing
the correct solution (i.e., $S_{S}$). In a next step, we establish the
\textit{distance} for transforming an empty model (i.e., $S_0$) to $S_{S}$ (i.e.,
the minimum number of change patterns needed). Typically, the process
designer has several possibilities to create a solution model $S_{S}$ by starting
from $S_0$ and applying a sequence of model transformations. From a cognitive
perspective, each possible sequence of change patterns that leads without detours
to the correct solution constitutes an \textit{optimal problem solving path}.
For example, inserting region R4 of Task B in Fig. \ref{img:tasksA+B}B starting
from an empty modeling canvas can be achieved in different ways with a minimum number of 4
change patterns (e.g., $S'$ can be created by first inserting \textit{R4.1}, next
embedding \textit{R4.1} in a conditional branch, then inserting \textit{R4.2},
and finally updating the transition condition).

\vspace{-7mm}  
\vspace{-2mm}

\begin{figure}[h!tb]
\vspace{-4mm}    
\center
  \includegraphics[width=0.8\textwidth]{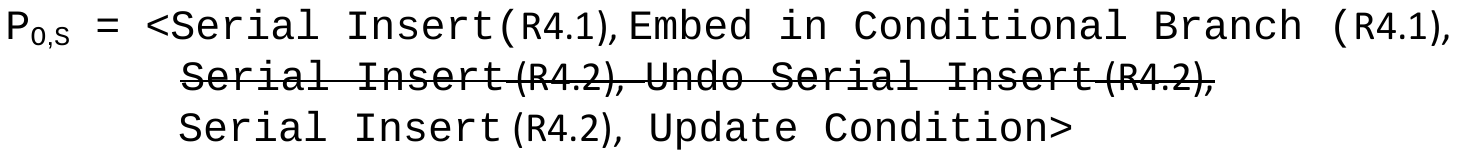}
  \caption{Problem Solving Path with 2 Process Deviations}
  \label{fig:deviation}
  \vspace{-6mm}
  \center
\end{figure}

\textit{Step 2: Determine Deviations from Optimal Problem Solving Path. } To
identify potential challenges designers were facing we
analyze their problem solving paths for both modeling tasks
using the replay functionality of CEP. For this, for each process designer we compare
the problem solving path $P_{0,S}$ (i.e., sequence of patterns to transform $S_0$
 to $S_S$) and capture deviations from the \textit{optimal problem solving path}.
Respective deviations can be detours the designer takes until coming up with the
correct solution. Deviations quantify how efficient the chosen problem solving
strategy is---denoted as \textit{process
deviations}. However, deviations can also be discrepancies between the model
created by designers and the solution model $S_S$, denoted as \textit{product
deviations}. Fig. \ref{fig:deviation} shows the problem solving path of one
process designer, who managed to model region R4 of Task B in
Fig. \ref{img:tasksA+B}B correctly (i.e., 0 product deviations), but made a
detour of 2 change patterns (crossed out lines) before reaching the solution (i.e., the
solution path $P_{0,S}$ comprises 2 superfluous change patterns summing up to 2
process deviations).
 
\textit{Step 3: Classification of Deviations and Aggregation of Deviations.}
In Step 3, using the replay functionality of CEP, deviations are mapped to
regions of the process model and reasons for every deviation are identified
(e.g., misinterpretation of the textual description, problems with usage of
patterns) in an iterative consensus-building process
\cite{DBLP:journals/is/ReckerSR12}. Moreover, to obtain an overview of which
model parts caused most difficulties we aggregate for each task deviations per
region.

\vspace{-6mm}

\subsection{Results Related to Task A}
\label{taskA}

Regarding Task A, a minimum of 18 operations is needed to create the correct
solution model (cf. Fig. \ref{img:tasksA+B}A). Overall, we identified 254
deviations---232 process deviations (i.e., detours in the modeling process) and
22 product deviations (i.e., deviations of the final models from the solution
model) (cf. Table \ref{numberDeviations}). Process deviations per process
designer ranged from 0 to 58, with an average of 13.4 deviations. In turn, product deviations ranged from 0 to 8 per process designer, with an average
of 1.3 deviations. 

\begin{table}[h!tb]
\vspace{-4mm}    
\center
  \includegraphics[width=0.8\textwidth]{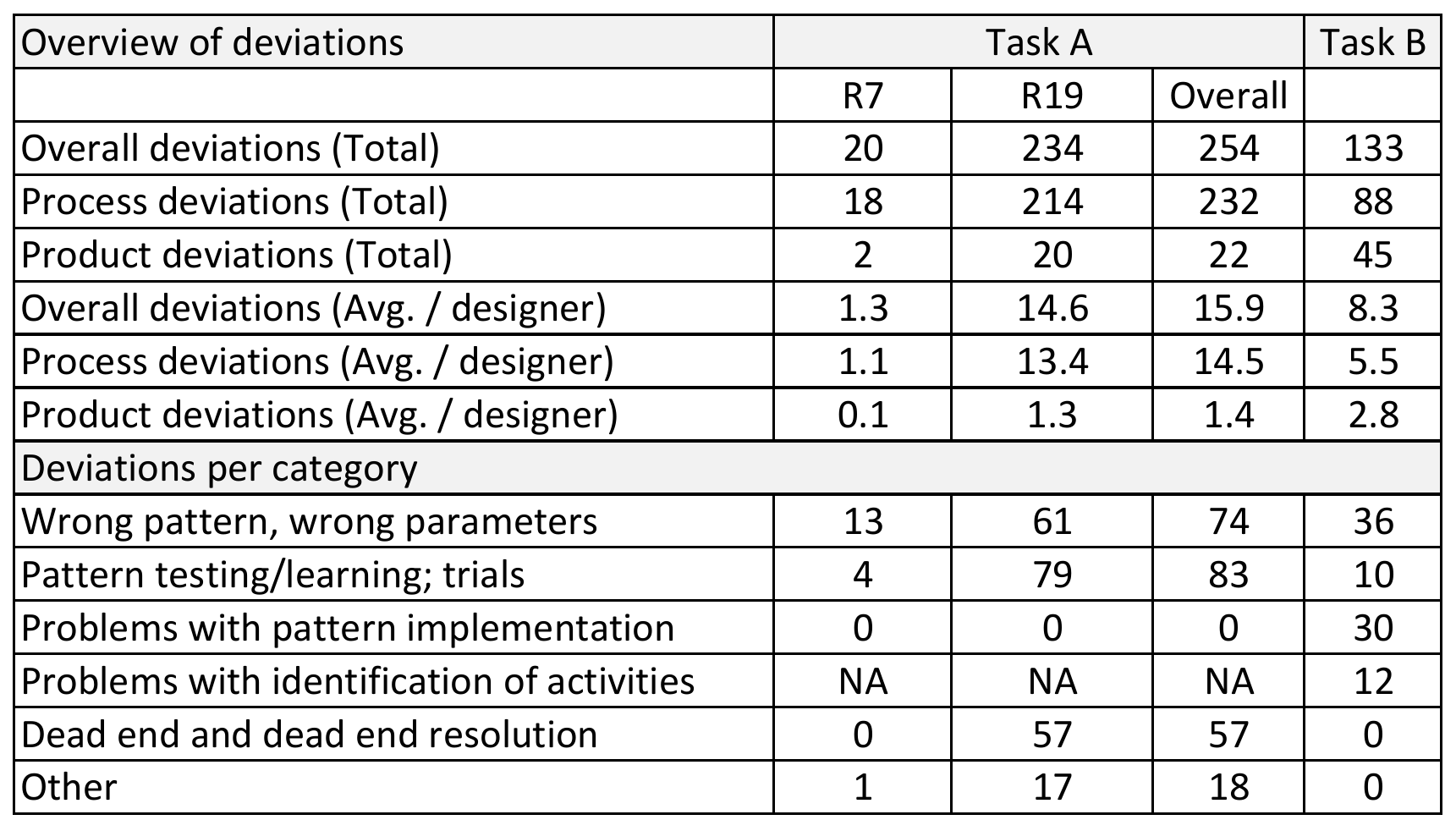}
  \caption{Overview of Results for Task A and Task B}
  \label{numberDeviations}
  \vspace{-6mm}
  \center 
\end{table}

\vspace{-2mm}

Most of the designers started the modeling with little problems (i.e., from the
232 process deviations only 18 are related to region R7) that were mostly caused
by wrong pattern usage (e.g., loop instead of conditional) or wrong parameter
settings (e.g., activity inserted at wrong position). The remaining 214
deviations occurred in region R19, which only comprises two activities more than
R7, but has higher structural complexity. Interestingly, when combining the
patterns in an optimal way, the creation of both parts requires a similar number
of change patterns (i.e., 7 patterns for R7 and 11 patterns for R19). When
analyzing the process deviations related to R19, it turned out that process
designers especially faced difficulties in creating region R16. Its nested
structure forces designers to apply patterns in a certain ordering (i.e.,
requires an effective problem solving strategy including look ahead). For
example, assuming that R8, R9, and R10 have already been inserted, the
insertion of R13 in parallel to the already inserted regions leads to a dead end
(i.e., the solution model can only be reached by deleting already created
parts). Most designers (11 out of 16) ended up at least once in a dead end when
trying to create the solution model. 3 of these 11 designers did not try to
resolve the dead end, i.e., R16 is modeled incorrectly in their final models.
The remaining designers (8 out of 11) tried to resolve the dead end by
backtracking in the modeling process (i.e., 57 deviations). Partially, it took them several
trials until they found a problem solving strategy suitable for constructing the
respective fragment (i.e., 79 deviations). Strategies for constructing R16
included the usage of dummy activities and experiments to test and learn the
functioning of the patterns. In turn, 5 out of 16 designers faced relatively
little difficulties with the creation of this fragment since their initial strategy
turned out to be effective, i.e., they were able to build the solution model in
a straight-forward manner. In addition to problems related to the creation of
R16, process deviations were caused in the context of single activities of
region R19. Again, the usage of wrong patterns or wrong parameter values was the
primary source of deviations (i.e., 61 deviations).

\subsection{Results Related to Task B}
\label{taskB}

Regarding Task B, a minimum of 19 change operations is needed to create the
correct solution model. Overall, we identified 133 deviations, 88 process
deviations and 45 product deviations (cf. Table \ref{numberDeviations}).
Process deviations per process designer ranged from 1 to 17 with an average of 5.5 deviations. Product deviations per
designer ranged from 0 to 5 deviations with an average of 2.8 deviations.
Product deviations were partially caused by ambiguities in the textual
description, partially by mismatches between the textual description and the
final models and presumably rather stem from problems with the domain than from
actual pattern usage. Since subjects did not have a process model given as a
template like in Task A, but had to build the model themselves starting from an
informal requirements description, it is little surprising that the percentage
of product deviations is much higher compared to Task A. Regarding process
deviations, for 12 out of 88, there is clear evidence that
they stem from problems with the domain rather than from problems caused by pattern usage (i.e., subjects were not sure whether to
include certain model parts as activities and presumably lacked schemata for
information extraction). For 30 out of 88 process deviations, there are clear
indications that they were caused by problems in pattern usage. As detailed
later on, all these deviations occurred in the context of region R4 and are the
result of tooling problems, i.e., caused by patterns implementation. Additional
10 deviations seem to stem from problems in pattern usage and were caused by
process designers testing the functioning of certain patterns. In turn, the
remaining 36 deviations were caused by process designers initially selecting the
wrong patterns (e.g., embedding an activity in a loop instead of embedding it in
a conditional branch), and by using the correct pattern with incorrect
parameters (e.g., inserting the correct activity at a wrong position). Even though most of
these deviations rather seem to be domain related, we cannot conclude with
certainty from our data whether this indicates problems caused by pattern usage
(i.e., mapping the mental model to the available patterns) or problems with the
domain (i.e., incorrect mental model).

\begin{figure}[h!tb]
\center 
  \includegraphics[width=0.4\textwidth]{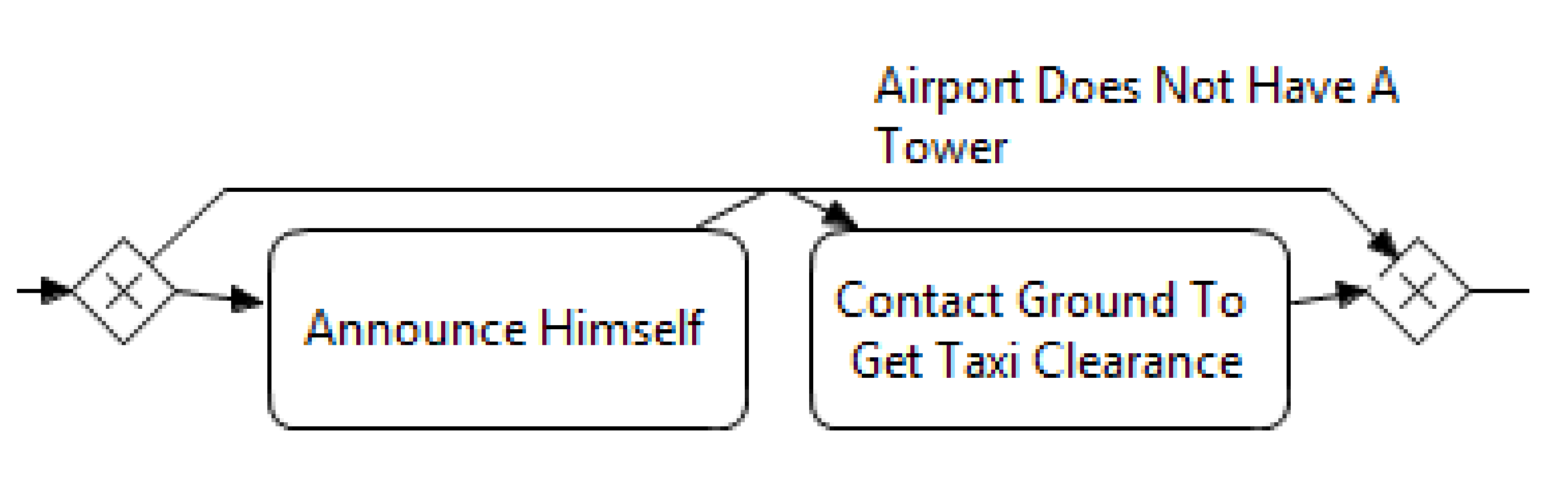}
  \caption{Problem with Combined Pattern Usage}
  \label{def:patternProblem} 
  \center
\end{figure}

Overall, subjects faced relatively little problems related to the usage of
patterns when working on Task B. In particular, they did not have any notable
problem when inserting activities in a sequence, making an activity optional,
or inserting an activity in parallel to another one. The only exception was a region
with two exclusive branches, which caused significant problems (cf. R4 in
Fig.~\ref{def:patternProblem}). To model this region correctly, process
designers had to first insert one of the activities using pattern \textit{Serial
Insert}, subsequently use pattern \textit{Embed Process Fragment in Conditional
Branch} to make the previously inserted activity optional, then use pattern
\textit{Serial Insert} to insert the second activity, and finally use pattern
\textit{Update Condition} to insert a transition condition. Even though the
subjects started correctly in the construction of this region---problems with the
automatic layout---CEP made 5 out of 16 process designers think that the second
\textit{Serial Insert} pattern has not been applied correctly resulting into
partially long detours (cf. Fig. \ref{def:patternProblem}). As a consequence,
they tried to apply the pattern several times even though their initial solution
would have been correct. 4 of the 5 process designers facing this problem
realized after a few trials that the pattern had been applied correctly. Just 1
of 5 process designers, however, created a workaround solution that correctly
reflected the requirements, but was a bit more complicated than the optimal
solution (i.e., instead of creating one process fragment with two conditional
branches, this designer created two process fragments with one optional activity
each). Overall 30 out of 88 process deviations were caused by this problem.
Interestingly, when faced with the same modeling structure later in the modeling
process again (i.e., R7), the process designers did not have these problems
anymore, but apparently learned how to use the patterns in combination to model
such construct (4 process designers) or how to circumvent the situation with a
workaround (1 process designer). While process designers had no problem solving
strategy available when constructing R4, they could rely on the plan schemata
developed in R4 for the construction of R7.

\vspace{-6mm}    
\vspace{-10mm} 

\vspace{-3mm}   
  
\section{Subjective Perception of Model Creation}
\label{easeOfUse}

This section addresses research question R2 which deals with the subjective
perception of process designers using change patterns. We investigate their ease
of use and perceived usefulness for creating process models and discuss feedback
provided by the participants after the exploratory study.
\subsection{Perceived Ease of Use and Perceived Usefulness} \label{easeOfUseAndPerceivedUsefulness} 
  
To assess in how far process designers with moderate process modeling knowledge
consider the CEP change pattern modeler as easy to use and useful, we asked them
to fill out the \textit{Perceived Ease of Use} and the \textit{Perceived
Usefulness} scales from the Technology Acceptance Model (TAM) \cite{Davis89}
after the modeling session. Both scales consist of six 7-point Likert items, ranging
from "Extremely likely" (1) over "Neither Likely nor Unlikely" (4) to "Extremely
Unlikely" (7). On average, for the Perceived Ease of Use scale the process
designer responded with 2.88, which approximately relates to "Slightly Likely"
(3).  For the Perceived Usefulness scale, in turn, the process designer in
average responded with 3.49, which approximately relates to "Slightly Likely"
(3). Hence, we conclude that process designers find it in average "slightly
likely" that it would be easy to learn and use change patterns.

\subsection{Qualitative Feedback Regarding Change Pattern Usage} 
\label{qualitativeFeedback} 
  
We additionally asked participants for qualitative feedback with the usage of
change patterns. The obtained feedback revealed usability issues, which are in line with the results reported
in Sect. \ref{taskA} and \ref{taskB}, and which at least partially explain why
perceived ease of use and perceived usefulness did not receive better scores. The
qualitative evaluation shows that perceived ease of use and perceived usefulness
of the patterns heavily depends on process characteristics. Several designers
stated that they perceive change patterns especially useful for models that are
rather simple, since in this case change patterns allow them to speed up
modeling. For more complex models (e.g., highly nested models), process designers
rather prefer using change primitives, since this gives them more flexibility:
\textit{``I like both ways of working, with patterns and without. After getting a
bit used to the patterns I find them easier to not let me make mistakes, for
thinking and sketching I find the other approach easier."} Since the available
pattern set was limited, process designers were partially forced to delete big
fragments when their problem solving strategy turned out to be ineffective (cf.
Sect. \ref{taskA} and \ref{taskB}). This was especially true for model regions
with complex structure requiring more sophisticated problem solving strategies on
how to combine patterns. This is reflected by the following statement of one
participant: \textit{``From my point of view, it always depends on whether
the process is clear, which makes it easy to use change patterns. However, if the
process somehow is really complex and it's hard to think about everything before
starting to model a process it's better to avoid using change patterns, since
once you've model something wrong, it could happen that you have to remodel many
parts of the process."} This statement indicates that limitations of working
memory might be a bottleneck, especially when designers lack schemata for
efficiently extracting and processing information and problem solving strategies
on how to best combine patterns. To combine the strength of patterns-based
modeling and the modeling of change primitives, designers expressed the wish for
a modeling environment allowing for the combination of both modeling
styles:\textit{``If the change pattern functionality would be included into
standard modeling tools it would be a very useful addition".}

%% file: discussion.tex
\vspace{-6mm}

\section{Discussion and Limitations}
\label{discuss} 

This section discusses the results and presents limitations that pose potential
threats to their validity.

 
\subsection{Discussion} 
\label{discussion}  

The results described in Sect. \ref{challenges} reveal that simple control flow
structures without any nesting can be well managed by most designers. Presumably,
designers were able to quickly develop plan schemata for simple models. In
general, only few problems, which can be directly attributed to pattern usage,
could be observed for model parts without any nesting. Even when faced with the
modeling environment for the first time, subjects did not have any notable
problems when inserting activities in sequences, making an activity optional, or
inserting an activity in parallel. There was only one exception where detours in
the modeling process were caused by a poor implementation of the graphical layout
of the CEP modeler, which will be addressed in a future version of this software
to improve perceived ease of use. Faced with more complex control flow
structures, in turn, the structural restrictions imposed by modeling based on
change patterns led to considerable problems with model construction partially
resulting into long detours or incorrect models. These findings are underlined by
feedback of the participants who appreciate the correctness-by-construction
guarantees, but feel restricted when faced with complex control flow constructs
(cf. Sect. \ref{easeOfUse}). Clearly, designers could not rely on existing plan
schemata for such complex structures, forcing designers to apply means-ends
analysis for solving the problem.

Difficulties faced by process designers can partially be explained by the
available patterns set. Even though the patterns available to process
designers cover all basic control-flow patterns (i.e., sequence,
exclusive/parallel branchings, and loops), the pattern set we used turned out to
be insufficient for efficient model construction. Especially in the context of
Task A most process designers had to delete parts of their model due to dead
ends when trying to construct region R16. Having a pattern \textit{Move Process
Fragment}~\cite{WRR08} available would presumably address many of the challenges
faced by process designers and facilitate resolution of dead ends. Therefore, we
plan to extend CEP and to conduct another study to test whether this will lead to
the expected benefits and improve perceived ease of use and usefulness
of change patterns modeling.


\subsection{Limitations} 
\label{limitations}

As every research, this work is subject to limitations. The fact that the sample
size (16 participants) was relatively small certainly constitutes a threat
regarding the generalization of our results. In addition, using students instead
of professionals poses another validity threat. However, we are mildly optimistic
about the usefulness of the presented insights on the basis of modeling behavior
of graduate students, since~\cite{ieee10-reijers} identified that such subjects
perform equally well in process modeling tasks as some professional designers.
However, we acknowledge that process designers experienced with the usage of
change patterns will presumably face less problems during model creation. Another
limitation relates to the fact that we used only two different modeling tasks
(with different complexity) in our study. The analysis indicated that
difficulties during model creation strongly depend on model characteristics. It
is questionable in how far results may be generalized to models with different
characteristics. As a consequence, we plan additional experiments testing the
influence of model structure on difficulties in change pattern usage. For some of
the deviations in the context of Task B we cannot conclude with certainty from
our data whether the problems were caused by pattern usage
or insufficient domain knowledge. To single out these factors we will conduct
further studies with a setup as suggested in \cite{Sof+12}. Regarding the
internal validity, to alleviate the thread related to the classification of
reasons for deviations, a consensus-building
process~\cite{DBLP:journals/is/ReckerSR12} was performed by two authors of the
paper.


%% file: relatedwork.tex
\section{Related Work}
\label{relatedWork}

%

Our work is closely related to research on the PPM and process model creation
patterns. Research on the process of modeling typically deals with 
interaction of different parties focusing on structured discussions among
system analysts and domain experts~\cite{FrWe06,HoPW05}. The procedure of
developing process models in a team is analyzed in~\cite{Ritt07} and
characterized as negotiation process. Participative modeling is discussed in
\cite{StPS07}. Each of these works builds on observations of modeling practice
and distills normative procedures for steering the process of modeling toward
successful completion. Hereby, the focus is on the effective interaction between
the involved stakeholders. Our work is complimentary to this perspective through
its focus on the \emph{formalization} of the process model. The interactions with
the modeling environment have been investigated in~\cite{PSZ+12}, identifying
three distinct modeling styles. In turn, \cite{CVR+12} demonstrates that a
structured modeling style leads to models of better quality,
and~\cite{PSZ+12,CVP+13} suggest different visualization techniques for obtaining
an overview of the PPM. \cite{PFM+13} investigates the PPM using eye movement
analysis. Common to all these works is the focus on interactions with the
modeling environment using change primitives, while this paper investigates the
use of change patterns. Also related to our work is the usage of change patterns
for process schema creation. For example, AristaFlow allows modeling a sound
process schema based on an extensible set of change patterns~\cite{DaRe09}. In
turn,~\cite{Gschwind08} describes a set of pattern compounds, similar to change
patterns, allowing for the context-sensitive selection and composition of
workflow patterns during process modeling. Complementary to existing research on
process model creation based on patterns, which has a strong design focus, this
paper provides first empirical insights into the usage of change patterns.


%% file: summary.tex
\section{Summary}
\label{summary} 

While work related to the PPM has emerged as a new stream of research in recent
years, little is known about this process when utilizing change patterns. In this
exploratory study we investigate the challenges, process designers are facing
when creating process models based on change patterns as well as their subjective
perception regarding the usage of these patterns. Our results show that process
designers face relatively little difficulties when creating simple control-flow
structures. When faced with more complex process structures, the structural
restrictions imposed by change patterns caused considerable problems for most of
the process designers. Building respective structures efficiently (i.e., without
detours) requires process designers to look ahead, since patterns cannot always
be arbitrarily combined. This need for looking ahead is a fundamental difference
compared to process model creation using change primitives and did not only lead
to observable difficulties, but was also perceived as challenging and restrictive
by subjects. The exploratory study not only confirmed that the creation of
process models using change patterns impacts the PPM, but also gives advice
regarding the improvement of the modeling approach based on change patterns. In
particular, it showed that a basic set of change pattern as used for the
exploratory study is not sufficient for efficient model creation.